# R2U++: A Multiscale Recurrent Residual U-Net with Dense Skip Connections for Medical Image Segmentation

Mehreen Mubashar[1], Hazrat Ali[1,2*], *Senior Member IEEE,* Christer Grönlund[2], and Shoaib Azmat[1]

[1] Department of Electrical and Computer Engineering, COMSATS University Islamabad, Abbottabad Campus, Abbottabad, Pakistan (Email: mehreenmubashar01@gmail.com, hazratali@cuiatd.edu.pk, shoaibazmat@cuiatd.edu.pk)
[2] Department of Radiation Sciences, Umeå University, Umeå, Sweden. (Email: christer.gronlund@umu.se)

*Abstract*— U-Net is a widely adopted neural network in the domain of medical image segmentation. Despite its quick embracement by the medical imaging community, its performance suffers on complicated datasets. The problem can be ascribed to its simple feature extracting blocks: encoder/decoder, and the semantic gap between encoder and decoder. Variants of U-Net (such as R2U-Net) have been proposed to address the problem of simple feature extracting blocks by making the network deeper, but it does not deal with the semantic gap problem. On the other hand, another variant UNET++ deals with the semantic gap problem by introducing dense skip connections but has simple feature extraction blocks. To overcome these issues, we propose a new U-Net based medical image segmentation architecture R2U++. In the proposed architecture, the adapted changes from vanilla U-Net are: (1) the plain convolutional backbone is replaced by a deeper recurrent residual convolution block. The increased field of view with these blocks aids in extracting crucial features for segmentation which is proven by improvement in the overall performance of the network. (2) The semantic gap between encoder and decoder is reduced by dense skip pathways. These pathways accumulate features coming from multiple scales and apply concatenation accordingly. The modified architecture has embedded multi-depth models, and an ensemble of outputs taken from varying depths improves the performance on foreground objects appearing at various scales in the images. The performance of R2U++ is evaluated on four distinct medical imaging modalities: electron microscopy (EM), X-rays, fundus, and computed tomography (CT). The average gain achieved in IoU score is 1.5±0.37% and in dice score is 0.9 ± 0.33% over UNET++, whereas, 4.21±2.72 in IoU and 3.47±1.89 in dice score over R2U-Net across different medical imaging segmentation datasets.

## I. Introduction

Image processing techniques have been applied to examine biomedical images for decades, and even to this day, designing computer-aided diagnostic systems (CAD) is one of the hot research areas [1]. The purpose of CADs is to design systems that can perform an accurate diagnosis of the underlying disease quickly, which can eventually aid in the treatment of a large number of patients. Quick diagnosis of diseases has shown a considerable decline in death rate, for example, in certain kinds of cancer tumors like brain tumors, kidney stones, stomach cancer, lung cancer, and breast cancer [2]. In this regard, a substantial amount of research effort has been put in this area with the target to improve and aid the processes of disease diagnosis from medical imagery.

The laborious nature of manual segmentation has increased the demand for automatic segmentation. Example images with segmentation masks are shown in Fig. 1. The traditional methods for CAD mostly based on hand-crafted features [3], [4] are now being replaced by variants of convolutional neural networks (CNN) models, such as AlexNet [5], VGGNet [6], and GoogleNet [7]. The proven success of CNNs over traditional methods has led to new variants of these techniques such as encoder-decoder architectures and deep generative models for different medical imaging applications [8], [9].

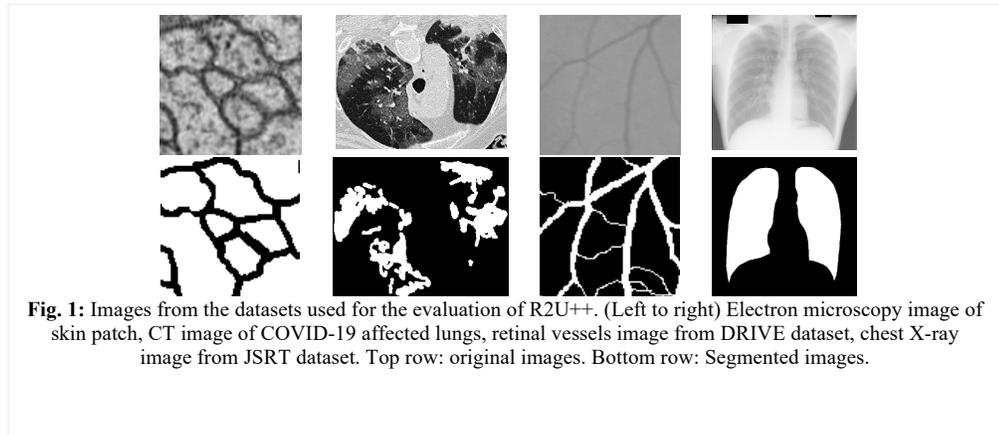

**Fig. 1:** Images from the datasets used for the evaluation of R2U++. (Left to right) Electron microscopy image of skin patch, CT image of COVID-19 affected lungs, retinal vessels image from DRIVE dataset, chest X-ray image from JSRT dataset. Top row: original images. Bottom row: Segmented images.

From the architectural standpoint, the models used for classification have a slightly different architecture than the ones used for segmentation. The classification models use an encoder and generate class probabilities as an output. On the contrary, as the segmentation demands capturing the context of an image alongside its location, it is crucial to have both encoding and decoding units in a network. The segmentation tasks in medical imaging, in general, are more sensitive and require extra refinement compared to natural images due to the associated healthcare decision-making. For example, the slight speculation around a lung nodule in a CT image is an indication of it being malignant; and its elimination from generated segmentation label would result in wrong clinical diagnosis. Therefore, there is always a need for improvement in segmentation models, so that they can correctly segment all the fine details of the object of interest.

The most adopted encoder-decoder structures in this regard are fully convolutional networks (FCN) [10] and the U-Net [11]. These two commonly used architectures differ in the way the skip connections help to retrieve the lost fine details. In FCN, the skip connections are used to sum up features of encoders with up-sampled decoder feature maps, while U-Net applies concatenation operation on these features. U-Net was the first medical imaging segmentation model shown in Fig. 2 (a) that outperformed all the models on small size medical imaging datasets. Due to U-Net simple architecture with plain convolutions in encoder/decoder, it becomes less efficient for some complicated medical imaging tasks [12], [13], [14], [15].

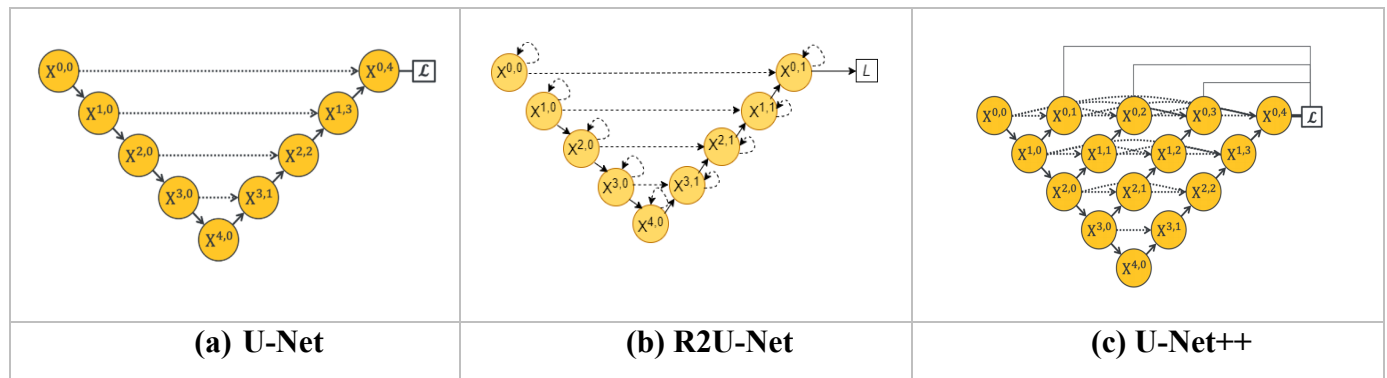

**Fig. 2: Overview of the architectures** (a) U-Net architecture with encoder and decoder blocks with simple skip connections (b) R2U-Net with recurrent residual convolution block and simple skip connections (c) U-Net++ with simple convolution blocks and dense skip connections.

In U-Net, the skip connections used between encoder and decoder require the concatenation to be at the same level. However, this concatenation, despite being at the same level, is not semantically similar [13], [15]. Therefore, several variants of U-Net have been proposed, with some attempting to change the backbone [16], [17] while others tweaking the skip connections between encoder and decoder [13], [15], [18]. The success of these variants to correctly classify the target objects in complex datasets can be attributed to two things: encoder/decoder blocks and skip connections [11], [13], [15]. The efficiency of the blocks being used as encoder/decoder enables the network to extract the features crucial for segmentation tasks. On the other hand, the skip/shortcut connections residing in between encoder and decoder help to recover the lost fine details of foreground objects. Considering the importance of these two factors, we have proposed an architecture that can enjoy the best of both worlds i.e., an efficient backbone and improved skip pathways. First, to focus on better feature accumulation, we have replaced plain convolution blocks of U-Net with recurrent residual convolution units adopted from [16] shown in Fig. 2(b). These recurrent units unfold to a predefined time step *t* making the network deeper at each layer. This increases the field of view in the lower layers of the neural network enabling them to extract precise low-level features. As the low-level features: the boundary of certain tumors, lungs, size of infection, are of utmost importance for the prognosis of the underlying disease; hence an accurate extraction helps to boost the network's performance. Second, the skip connections of vanilla U-Net have been replaced by dense skip connections adopted from U-Net++ [13], [15] shown in Fig 2(c). In vanilla U-Net feature maps coming from the encoder are at a lower level than the feature maps of the decoder, this semantic difference is called semantic gap. These dense skip connections reduce the semantic gap between encoder and decoder features before concatenation. Besides, these dense connections are forwarding the different scale information to the decoder. The decoder can then perform the aggregation on various scale features to enhance the segmentation accuracy. These architectural modifications have introduced multi-depth embedded models partially sharing a common encoder. In addition, training the network under deep supervision performs shared learning on all the embedded depths which is highly beneficial for segmenting multiscale foreground objects. Our main contributions are:

1. We introduce a new deeper segmentation model namely R2U++ for medical image segmentation. The model uses recurrent residual blocks over vanilla convolutional blocks which provide a large field of view even in the lower layers to extract features enriched with lower-level information. As we replace the plain convolution blocks of U-Net with recurrent residual convolution units, these recurrent units unfold to a predefined time step t making the network deeper at each layer.
2. We use dense skip pathways. The dense skip pathways reduce the semantic gap of the concatenating encoder and decoder and propagate different scale information to the decoder. The dense skip connections also improve the gradient flow.
3. The concept of dense skip pathways also enables us to define an architecture where multiple architectures of different depths are merged into a single architecture. The ensemble of multi-depth can capture the information of varying size objects.
4. Equipped with the above characteristics, our resultant residual recurrent architecture with dense skip connections has consistently outperformed the existing models on medical images of different modalities including electron microscopy (EM) images of skin lesions, computed tomography (CT) images of COVID-19 affected lungs, Chest X-Ray images, and retinal fundoscopic images of retinal vessels.

The remainder of this paper is organized as follows. In Section II, we discuss the related work. The proposed architecture is explained in Section III. The datasets used in the study and the experimental details are presented in Section IV. Results are presented in Section V. The paper is concluded in Section VI.

## II. Related Work

Semantic segmentation refers to the kind of labeling where we have to assign a label to each pixel of an image. In the domain of segmentation, the work on fully convolutional neural networks (FCN) introduced the concept of combining what and where information to properly label the pixels of an image [10] . It was achieved by adding a link between the coarse and the fine layers. In [19], Chen et al., proposed *deeplab* for semantic image segmentation using atrous convolution, which not only increased the field of view but atrous spatial pyramid pooling (ASPP) enabled them to segment objects at multiple scales. SegNet [20] is a corresponding encoder-decoder segmentation network, in which the encoder is similar to the VGG network [6] with no fully connected layers at the end. However, its major contribution was the use of max pooling indices in decoder layers from its corresponding encoder part. Most of these architectures use large data and are designed specifically for computer vision applications. The major problem that initially hampered the success of convolutional neural networks in the domain of medical image segmentation was the unavailability of sufficient medical images for training deep models. However, this problem was first of all tackled by the segmentation network U-Net [11], specifically designed for medical image segmentation tasks and worked relatively well even for smaller datasets. Since then, U-Net has become a popular choice for medical image segmentation tasks.

The U-Net is built upon FCN [10], which comprises two paths: the contracting path and the expanding path. The contracting path has a traditional convolutional encoding unit that performs convolution operations followed by rectified linear units (ReLU) activation. It is then down-sampled via 2×2 max pooling. The main modification of this architecture was to have a symmetric expanding path with a large number of feature channels obtained through up-convolution. In the expanding path, up-sampling is followed by up-convolution, which reduces the number of feature maps to half. These features are then concatenated with the feature maps from the corresponding encoding unit. The architecture was adopted quickly due to its several advantages. Firstly, it captures context and location information simultaneously. Secondly, it meets the demand for a network that can provide better results on small medical imaging datasets. Finally, it is trained in an end-to-end fashion and provides a segmentation mask in the forward pass. Nonetheless, U-Net is not restricted to medical imaging only but has also proven its efficiency in many computer vision applications [21]. Several variants of U-Net have been proposed to adopt the simple U-Net architecture to complex datasets. These alterations can be broadly classified into two categories: changing the backbone and reforming the skip connections – as discussed below.

*A. Modified Backbone*

The U-Net model uses two convolution layers in each encoder-decoder block which makes it very simple for complex datasets. One of the ways adopted by researchers to deal with the problem is to increase the depth of the network. However, increasing the depth is not as easy as stacking layers. The networks with a depth of tens of layers initially faced the issue of vanishing gradients [22]. The issue has been addressed by using different activation functions like ReLU, Exponential Linear Units (ELU) [6], [7], and by applying normalization in between the layers [23]. In [24], He et al. pointed out the degradation problem: increasing the network's depth saturates the performance and then promptly drops it. To overcome this problem, they proposed the solution of using identity mapping or skip connections in their proposed Residual Network (ResNet). The ResNet learns via residual function and makes the optimization task easier. This approach helped with overcoming the degradation problem and improved the network's performance. Ever since, deep models and skip connections go hand in hand. These residual connections are quite popular in deep U-Net variants; like in [16], the authors have devised Recurrent Residual U-Net (R2U-Net). The model is a modification of U-Net [11] with replacing simple convolutional units with Recurrent Residual Convolutional Layers (RRCL) [25], [24]. Each encoder-decoder unit has two sub RRCNN blocks where each unfolds to a time step *t*. The final output is an element-wise summation of output from the second recurrent convolution block and the original input. The increased field of view even in the lower layers and the efficiency of feature summation aids in extracting very low-level features, which are crucial for medical image segmentation. This architecture with fewer parameters outperformed the ones with a large number of parameters. In [26], however, this element-wise feature summation did not benefit in improving the testing performance due to the summation being performed outside the network. Similarly, in M-UNet [27], the authors have made the network sufficiently deep by embedding DenseNet [28] in the architecture. The convolution blocks of the encoder are replaced by DenseNet, while the plain convolutions are kept in the decoder block. The arrangement has made the network deeper that improved performance while keeping a reasonable number of network parameters. DIU-Net [29] is an attempt to make the U-Net model wider and deeper by fusing Inception-Res and dense inception block. Unlike traditional Inception-Res block, each convolution layer is followed by a batch normalization layer to avoid vanishing gradient. The dense inception block comprises densely connected inception blocks. The network uses 3 dense inception blocks, with one in the encoder, one in the decoder, and one in the middle. The dense inception block of synthesis and analysis path has 12 inception blocks, whereas the middle one uses 24 blocks. Experimentation results showed improvement over state-of-the-art models. However, the downside of the network is that increasing the growth rate will lead to too many network parameters, which makes the training process slower and difficult. Likewise, in MultiResUNet [12], the encoder-decoder blocks are replaced by a MultiRes block which makes use of residual connections. The motivation behind MultiRes blocks is to make the network capable of segmenting the foreground objects appearing at various scales in medical images. These blocks implement Inception-like blocks [7] of 3x3, 5x5, and 7x7 with successive 3x3 filters and a 1x1 convolution added with residual connection to preserve the dimensionality of the image. The architecture has shown significant improvement in performance over U-Net across five medical image modalities. With the focus on extracting advanced segmentation features, probabilistic programming is used in [30] with U-Net to enhance performance on ultrasound nerve segmentation. Similarly, in the residual attention U-Net model [31], the authors have used aggregated residual transformation and soft attention in the decoder. The aggregated residual block made the network efficiently deep, which was highly crucial for extracting efficient features for a complex multi-class problem. The network outperformed the U-Net on segmentation of the COVID-19 dataset. Another encoder-decoder network presented in [32] proposes the residual block and feature variation (FV) unit. These two blocks are used in the first three layers of the encoder. In the fourth layer, progressive atrous spatial pyramid pooling is added to increase the receptive field. However, the decoder of the network comprises simple deconvolution blocks. The architecture demonstrates the importance of the increased receptive field in the performance of a model.

*B. Modified Skip Connections*

Most of the variants of U-Net, including those designed for targeting 3D medical images [34], [35], have been using the plain skip connection. The effectiveness of skip connections in recovering the lost fine-grained details has also been demonstrated in many other segmentation architectures like [38], [39], [40], [41] and has been proven by Drozdzal et al. in [42].

Zhou et al. [13], [15] brought attention towards redesigning the skip connection between the encoder and the decoder networks. In U-Net [11], the features from the encoder are directly concatenated with the decoder which requires that they are at the same scale. However, the authors in [13], [15] argued that even though these feature maps are at the same scale, but not semantically similar and there is no theory to back that this fusion is the best possible strategy. Therefore, they replaced these simple connections with dense convolutional blocks to enrich encoder features with semantic information and bring their semantic level closer to the awaiting decoder before merging. In this way, the optimization task becomes easier. Another contribution was to introduce an ensemble of U-Nets with different depths making the model capable of segmenting objects of varying sizes with high accuracy. These dense skip connections are quickly adopted by researchers in models for various applications such as gallstone segmentation [36], pelvic organ segmentation [37], and brain tumor segmentation [43], [14]. The use of these dense skip connections in [15] has proven efficacy in Mask-RCNN segmentation as well. Likewise, the Dense U-Net++ [14] uses Half Dense U-Net [33] with the dense skip connections along with the skip pathways. The dense block at each layer uses the aggregated features from all the previous layers. It highlights the benefit of combining the dense skip pathways with aggregated features from Half Dense U-Net. MDU-Net [18] redesigned the skip connections to connect each decoder with three encoders depending on their position. In addition to this, the network uses skip connections along each encoder-decoder block to connect it with all the previous blocks. These connections enable them to use features from different scales. The architecture demonstrates the importance of using the features from various scales with feature concatenation from a different encoder for gland segmentation. Different medical imaging segmentation models and variants of U-Net are summarized in Table I.

**Table I:** Development of Medical Imaging Segmentation Models over the years

| Reference | Model Type | Model Name | Key Feature |
| --- | --- | --- | --- |
| [10] | Baseline Encoder-Decoder Models of Computer Vision | Fully convolutional Neural Networks (FCN) | Encoder Decoder Introduction |
| [19] | | Deeplab | Atrous convolution |
| [20] | | SegNet | VGG like encoder |
| [11] | Medical Imaging Segmentation Model | U-Net | Designed specifically for medical images with U like encoder-decoder structure |
| [16] | U-Net Backbone Modification | R2U-Net | Recurrent Residual Convolutional Layers in U-Net |
| [27] | | M-UNet | Embedding dense layers in U-Net |
| [29] | | DIU-Net | Fusing Inception-Res and dense inception block in U-Net |
| [12] | | MultiResUNet | Introduction of Inception like blocks in U-Net |
| [31] | | Residual attention U-Net | Aggregated residual transformation and soft attention in the decoder of U-Net |
| [33] | | Half Dense U-Net | Hybrid dense convolution block with different filters |
| [34] | | 3D U-Net | 3D Convolutional Layers in U-Net |
| [35] | | V-Net | U-Net modification for volumetric medical Images |
| [13] | U-Net Skip Connections Modification | U-Net++ | Dense skip connections between encoder and decoder |
| [15] | | U-Net++ | Dense skip connections with modified backbones of existing architectures |
| [36] | | U-Next | Attention along decoder, spatial pyramid pooling of skip connection and dense connection along skip pathways |
| [37] | | Dilated Convolution U-NET++ | Dilated convolutions in backbone and along skip pathways |
| [14] | | DU++ | Backbone of Half Dense U-Net and dense skip pathways |
| [18] | | MDU-Net | Multiscale dense skip connections between encoder and decoder |

### III. Proposed Network Architecture: R2U++

To overcome the challenges of U-Net [11] and its variants as mentioned in Section II, we propose a model R2U++. The three main components for the proposed network, namely the skip pathways, the backbone, and the deep supervision, are described below.

*A. Skip pathways*

Re-designed skip pathways modify the connection between encoder and decoder. Inspired from U-NET++ [13], [15], the feature map coming from the encoder will go through dense skip pathways before entering into the decoder block. The dense skip pathways refer to the dense skip connections to the convolution blocks along the skip pathway. The number of convolution layers along the skip pathways is determined according to its pyramid level. As shown in Fig. 3(d), for example, if encoder and decoder are at level 4, encoder block is $X^{(0,0)}$ and decoder block is $X^{(0,4)}$, there will be three convolution blocks: $X^{(0,1)}, X^{(0,2)}$ and $X^{(0,3)}$ in the dense skip pathway. Each convolution layer along the skip pathway applies convolution on the concatenated feature maps coming from all the previous blocks at the same level and the corresponding up-sampled feature map from the lower block. For example, $X^{(0,2)}$ applies convolution on the concatenated feature maps coming from the same level blocks: $X^{(0,0)}$, $X^{(0,1)}$ and up-sampled feature map from lower block $X^{(1,1)}$. In this way, the multiscale features with the same resolution are combined horizontally, whereas different resolution multiscale features are combined vertically. It will not only reduce the feature gap between encoder and decoder but will also capture the multiscale context.

Mathematically, skip pathways can be formulated by equation (1). Let us assume $m$ to be the index of the down-sampling layer in the case of encoder, and $n$ to be the index of convolution layer residing in the skip pathways. The concatenated input to the convolutional layer $X^{(m,n)}$ can be expressed as:

$$x_i^{m,n} = \begin{cases} x^{m-1,n} & n=0 \\ [[x^{m,k}]_{k=0}^{n-1}, u(x^{m+1,n-1})] & n>0 \end{cases} \quad (1)$$

The feature map for the $X^{(m,n)}$ convolutional layer then becomes:

$$x_o^{m,n} = H(x_i^{m,n}) \quad (2)$$

Where $H(.)$ is the representation of recurrent residual convolution explained in III.B. The up-sampling from the lower level is denoted by $u(.)$. The concatenation operation is represented by large square brackets. It can be noticed from Fig. 3 that the outermost encoder with $n=0$, is fed with only one input from its upper encoder block. However, the encoders with $n=1$ receive two inputs; one from the same encoder level and one up-sampled input from the lower level of the encoder. Due to the dense skip connections, for the nodes with a value of $n>1$, $n$ inputs are received from the same encoder level, and one input is up-sampled from the lower corresponding encoder level.

**Table II:** Details of the architectures and number of filters used in each convolution block $X^{m,n}$. Comparison of number of parameters with different value of $t$

| Network | t | Params | $X^{0,0-4}$ | $X^{1,0-3}$ | $X^{2,0-2}$ | $X^{3,0-1}$ | $X^{4,0}$ |
|---|---|---|---|---|---|---|---|
| U-NET [11] | - | 7.0M | 32 | 64 | 128 | 256 | 512 |
| R2U-NET [16] | 2 | 16.7M | 32 | 64 | 128 | 256 | 512 |
| U-NET++ [15] | - | 9.0M | 32 | 64 | 128 | 256 | 512 |
| R2U++ (Ours) | 1 | 9.7M | 32 | 64 | 128 | 256 | 512 |
| R2U++ (Ours) | 2 | 18M | 32 | 64 | 128 | 256 | 512 |

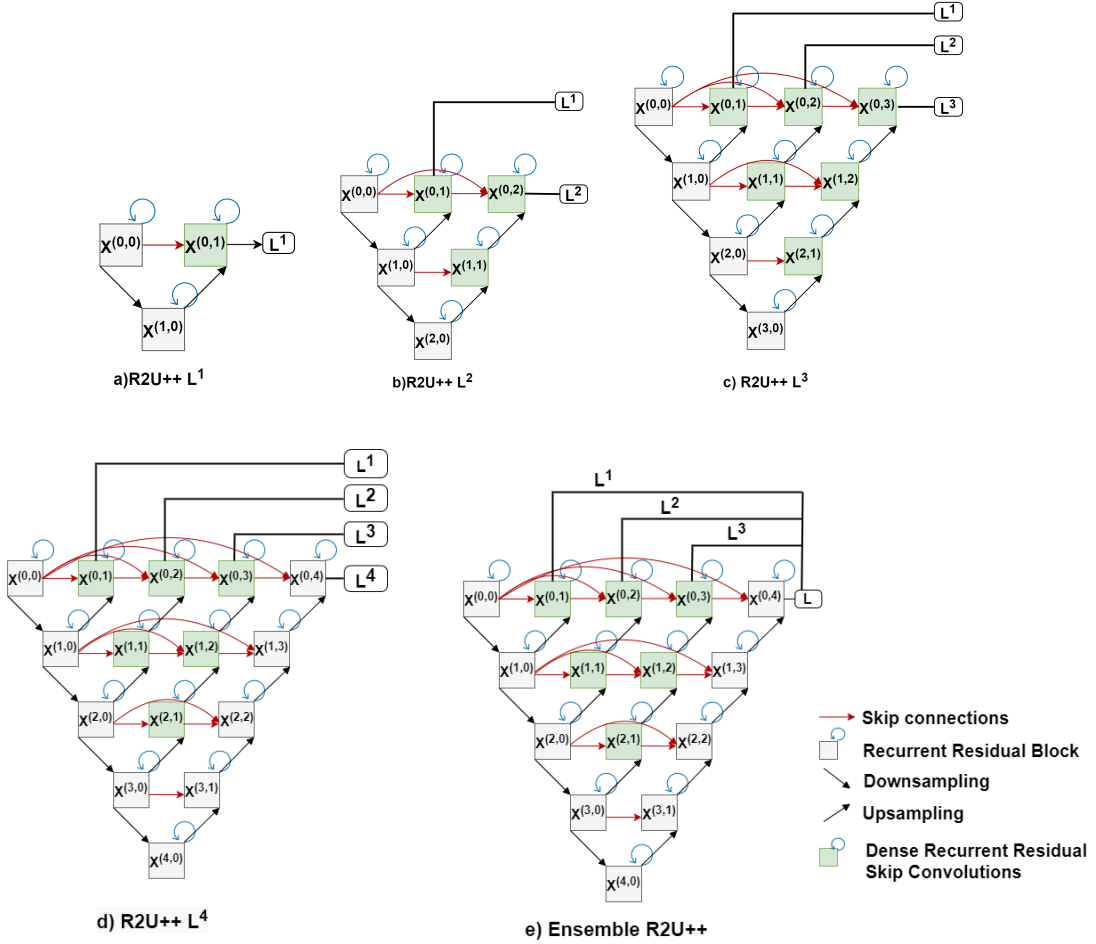

**Fig. 3:** R2U++ with evolving depths from (a) to (d). Ensemble of these depths is shown in (e). Each convolutional block performs recurrent convolution depending on the time step *t*, as shown for *t*=2 in Fig. 4. Residual connections are added between recurrent convolutions to avoid degradation problems (as shown in Fig. 4 later). (a) – (d) R2U++ with $L^{1-4}$ depths; every decoder in all the depths receives similar resolution multiscale features horizontally from its corresponding dense skip pathways, whereas varying resolution multiscale features are aggregated vertically across the network. (e) Average ensemble. In average ensemble network, all of these networks have their own decoder but partially share the same encoder which introduces shared learning in the network. R2U++ can explicitly benefit from deep supervision as depths like $L^2, L^3$ and $L^4$ are embedded with their corresponding lower-level networks.

*B. Backbone*

The U-Net model and its variants have been reporting leading results on several medical image segmentation datasets. Inspired by one of the variants, the Recurrent Residual U-Net [16], we have used recurrent residual convolutions layers (RRCL) over the simple convolutional layers of U-Net. The recurrent convolution layer (RCL) grows in accordance with time steps [25]. Let us define discrete time step as *t*. To represent the RRCL, we define the H(.) operation at time step t as RRCL. The feature map according to [16] can be represented as:

$$(O^{m,n})_t = (w^{m,n})_t^f * (x_i^{m,n})_t^f + (w^{m,n})_{t-1}^r * (x_i^{m,n})_{t-1}^r \qquad (3)$$

Here, the concatenated inputs for the RCL are expressed as $(x_i^{m,n})_t^f$ and $(x_i^{m,n})_{t-1}^r$ respectively. The term $(w^{m,n})_t^f$ represents the weights in a standard convolution operation, whereas $(w^{m,n})_{t-1}^r$ represents weights in a recurrent convolution operation. The output $(O^{m,n})_t$ generated from recurrent convolution block fed to the activation function ReLU which is represented as:

$$(O^{m,n})_t = f((O^{m,n})_t) = max(0,(O^{m,n})_t) \quad (4)$$

This output of the RCL unit at time step $t$ is then passed to the succeeding RCL unit of RRCL. If $(F^{m,n})_t$ is the output from the second RCL unit of RRCL then the final output from RRCL is computed as:

$$(x_o^{m,n})_t = (x_i^{m,n})_t + (F^{m,n})_t \quad (5)$$

Here, $(x_o^{m,n})_t$ shows the output of the RRCL unit at time step $t$. This output is then fed to the down-sampling layer in the case of the encoder, to the up-sampling layer in the case of the decoder, and to the next recurrent residual convolution layer (RRCL) in case of skip pathways.

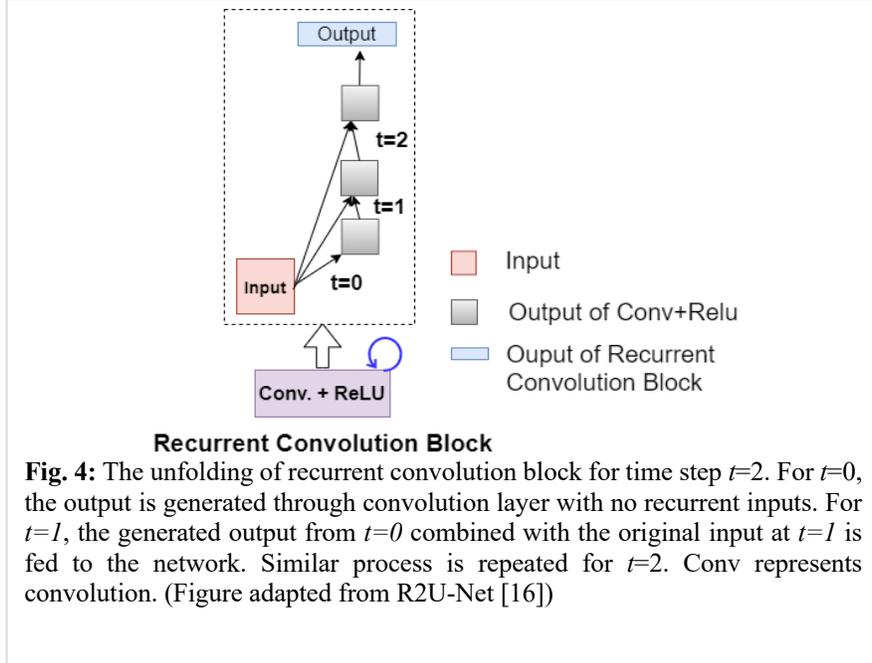

**Recurrent Convolution Block**
**Fig. 4:** The unfolding of recurrent convolution block for time step $t=2$. For $t=0$, the output is generated through convolution layer with no recurrent inputs. For $t=1$, the generated output from $t=0$ combined with the original input at $t=1$ is fed to the network. Similar process is repeated for $t=2$. Conv represents convolution. (Figure adapted from R2U-Net [16])

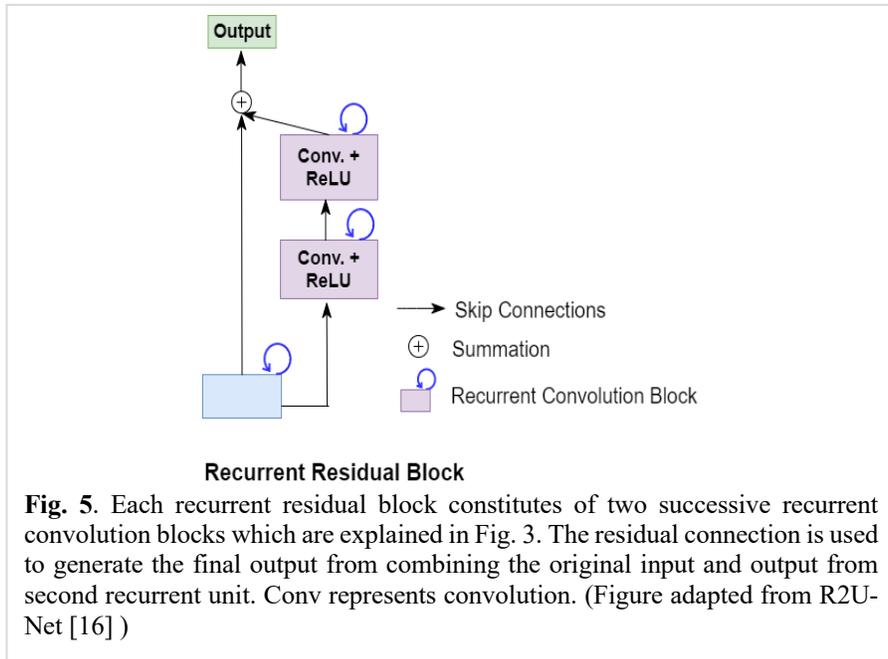

**Recurrent Residual Block**
**Fig. 5**. Each recurrent residual block constitutes of two successive recurrent convolution blocks which are explained in Fig. 3. The residual connection is used to generate the final output from combining the original input and output from second recurrent unit. Conv represents convolution. (Figure adapted from R2U-Net [16])

The visual representation of unfolding of RCL for *t=2* is shown in Fig. 4. For the convolution operation at *t=2*, the current input at *t=2* and the output from previous time step *t=1* both are applied with convolutional operation according to equation 3 and 4. Each recurrent residual block as shown in Fig. 5 further comprises two recurrent convolution blocks. The input sample when fed into the recurrent residual block passes from two back-to-back recurrent convolution blocks. The final output from recurrent residual block is the feature-wise summation of the original input at time step *t* and output from the second RCL block at time step *t*. All the convolutional blocks in R2U++ are recurrent residual convolution blocks.

*C. Deep Supervision*

The added dense skip connections enable the network to merge the architectures of various depths into a single architecture, as shown in Figure 3. Different depths are separately shown in Figure 3(a-d), where 3(a) shows the architecture with only one decoder making the architecture to be a level-1 network. However, level-2 architecture is shown in 3(b) with level-1 $X^{(0,0)}$, $X^{(0,1)}$ and $X^{(1,0)}$ embedded in it. Similarly, level-3 and level-4 are shown in 3(c) and 3(d). For 3 (a-d), the output is taken from $L^1, L^2, L^3$ and $L^4$, respectively. These networks are trained without deep supervision using equation 6. Fig. 3(e) refers to the ensemble network; when the final output is taken as an average of output from different depths. Ensemble architecture shown in Fig. 3(e) is a level 4 network embedded with all lower depths i.e. $L^1, L^2$ and $L^3$. All of these four levels share the same encoders but have their own decoders. Each of the levels is trained separately with its own loss function i.e., $X^{(0,q)}$ where $q\epsilon\{1,2,3,4\}$. At the inference, the final output will be calculated by taking the average of the output from each depth. It is trained using deep supervision scheme in R2U++, the loss function is applied on the nodes $X^{(0,q)}$ where $q\epsilon\{1,2,3,4\}$. A 1x1 convolution layer followed by activation function is added at the output of nodes $X^{(0,1)}, X^{(0,2)}, X^{(0,3)}$ and $X^{(0,4)}$. This convolution layer has C number of filters for the C segmentation classes in any dataset. We have used the loss function defined for the U-Net++ in [13], [15]. It is a hybrid loss function that comprises pixel-wise cross entropy loss and soft dice coefficient loss. The loss function is calculated for each of the semantic level i.e. $X^{(0,1)}, X^{(0,2)}, X^{(0,3)}$ and $X^{(0,4)}$. The hybrid loss function can enjoy the perks from both loss functions: smooth gradients and dealing with class imbalance problems. Mathematically, it can be written as:

$$L(Y,P) = -\frac{1}{N}\sum_{c=1}^{C}\sum_{n=1}^{N}\left(y_{n,c}\log p_{n,c} + \frac{2y_{n,c}p_{n,c}}{y_{n,c}^2+p_{n,c}^2}\right) \quad (6)$$

Where, Y denotes the ground truth labels, P denotes the predicted probabilities values, C represents the number of segmentation classes. Furthermore, $y_{n,c} \in Y$ and $p_{n,c} \in P$, where n denotes the $n^{th}$ pixel in a batch with a total of N pixels within a given batch. Finally, the total loss is the weighted sum of the individual loss functions. Mathematically, it can be written as:

$$L = \sum_{i=1}^{d}\eta_i \cdot L(Y,P^i) \quad (7)$$

The summation runs over the number of decoders represented by d. The value of $\eta_i$ is set to be one to assign the same weight to all the decoder losses.

To sum up the benefits of our architecture, the Residual Unit helps in training a deeper architecture by avoiding degradation problems. The Recurrent Unit aids in feature accumulation, which enables it to accumulate accurate low-level features highly crucial for segmentation. Having convolution layers on skip pathways reduces the dissimilarity in the features of the encoder and decoder. Having dense skip/shortcut connections on skip pathways improve gradient flow. Finally, ensemble multi-depth outputs ensure better accuracy on multiscale foreground objects.

IV. EXPERIMENTS

The experimentation process involves two main steps; training and testing, as shown in Fig. 6. For training, pre-processed images are fed to R2U++ to train the model using cross validation. Once the training process is completed, unseen testing data is presented to the trained model to make predictions.

*A. Datasets*

The proposed architecture has been evaluated on a range of biomedical image segmentation datasets, namely: (i) Electron Microscopy (EM) dataset of skin lesions, (ii) COVID-19 dataset of lung CT images, (iii) DRIVE dataset of retinal fundoscopic images, and (iv) JSRT dataset of chest X-ray images. These datasets cover the segmentation of skin lesions, lungs, and retinal blood vessels, as shown in Fig. 1. These datasets are generated from medical image modalities like microscopy, CT scans and X-rays.

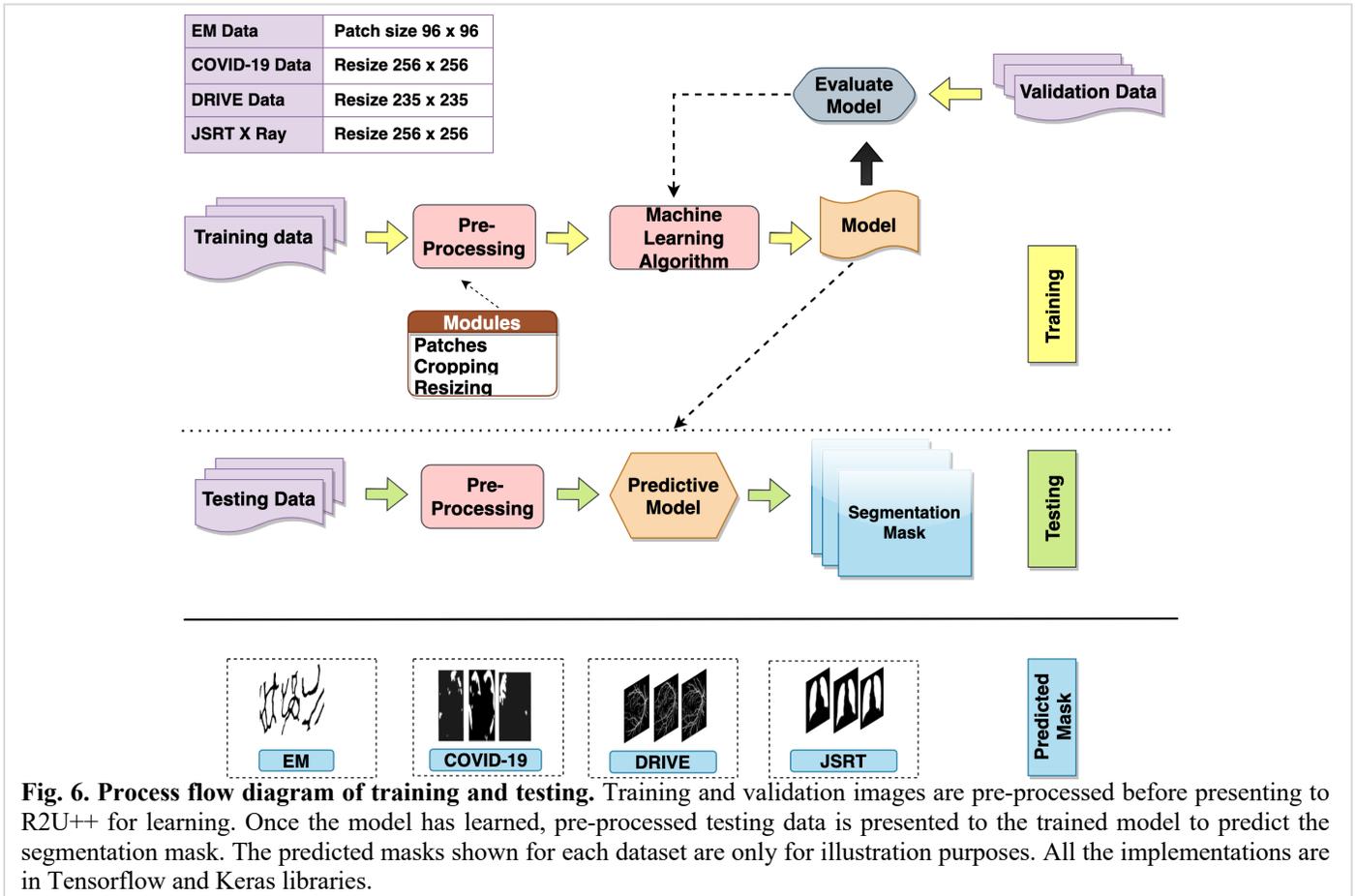

**Fig. 6. Process flow diagram of training and testing.** Training and validation images are pre-processed before presenting to R2U++ for learning. Once the model has learned, pre-processed testing data is presented to the trained model to predict the segmentation mask. The predicted masks shown for each dataset are only for illustration purposes. All the implementations are in Tensorflow and Keras libraries.

*1) Electron Microscopic (EM):* This publicly available dataset is a part of the ISBI 2012 EM segmentation challenge [44]. The dataset comprises a total of 30 images, with each having a dimensionality of 512×512. These images are extracted from serial section transmission electron microscopy (ssTEM) of the Drosophila first instar larva ventral nerve cord (VNC). The dataset is provided with the fully annotated ground truth labels for each image. The cells are labeled as white, whereas the membranes are represented with the black pixels. For the experimentation purpose, we randomly split the dataset into training 27 images from which 3 images are used for validation while testing is performed on the remaining 3 images. To overcome the small sample size of images, we have used the patch-based strategy for both training and inference. All the patches are generated using the sliding window technique with a patch size of 96×96 and a stride of 48.

*2) COVID-19 CT Images Dataset:* It is the first publicly available dataset for the COVID-19 segmentation [45]. The dataset comprises a total of 100 CT scans extracted from 19 COVID-19 patients. These images are gathered by the Italian Society of Medical and Interventional Radiology. The ground truths of only 100 slices are publicly available. To overcome the small sample size of labeled ground truth, another dataset is generated in [46] by extracting the unlabeled images from COVID-19 CT segmentation dataset. The unlabeled CT volumes from all 19 patients are extracted and pseudo labels for the 1600 2D slices from these volumes are generated. We have used these pseudo labels from [46] to pre-train our network. Subsequently, these weights are used to initialize our network. From 100 labeled slices, 45 randomly selected images are used for training, 5 for validation, and 50 images are used to evaluate the model's performance. As these images are not of uniform dimension, so we resized all images to 256×256.

*3) JSRT dataset of chest X-ray images:* The dataset used for lung image segmentation is produced by the Japanese Society of Radiological Technology (JSRT) [47]. The dataset contains 247 chest X-Rays with 154 nodule images and 93 non-nodule images The resolution of images is 2048×2048. We have split the dataset into 80% training and 20% testing. From training images, we have used 38 images for validation. We have resized the images to 256×256 to reduce the computational complexity.

*4) Blood Vessel Segmentation:* In our experimentation, we have used the DRIVE database [48] for retinal blood vessel segmentation. The dataset has in total 40 retinal images. The dataset is split into 20 training images and 20 testing images. Each image has a dimensionality of 565×584. In order to square the image dimensions, we have cropped images, taking the portion from 19 to 554 rows and 29 to 564 columns. The resulting images were of size 535×535. In our experimentations, we have used patch-based technique for both training and inference. The patches are extracted using the sliding window technique with a patch size of 96×96 and stride 5. We have generated 154880 testing and 154880 training patches from which 30976 are used for validation.

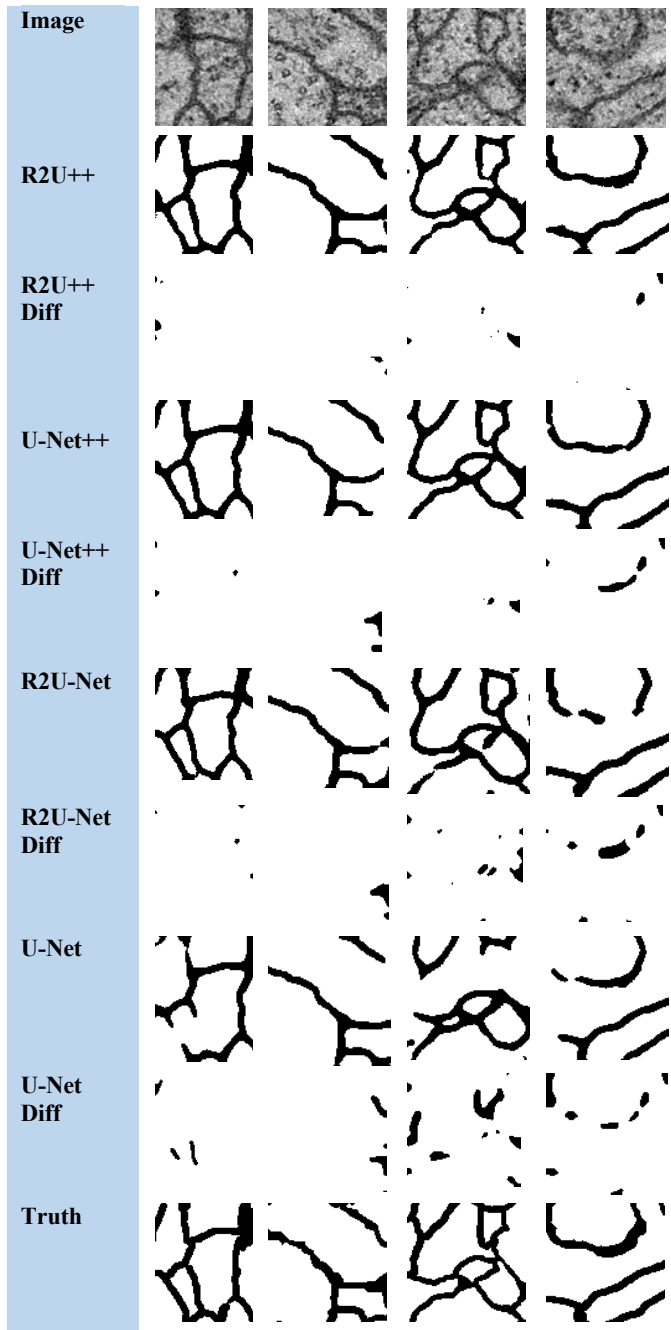

**Fig. 7:** The semantic segmentation outputs and difference images with ground truth for EM dataset from R2U++ (Ours), U-Net++, R2U-Net, and U-Net. The first row has the input image, and the final row contains the ground truth image. Diff represents the difference.

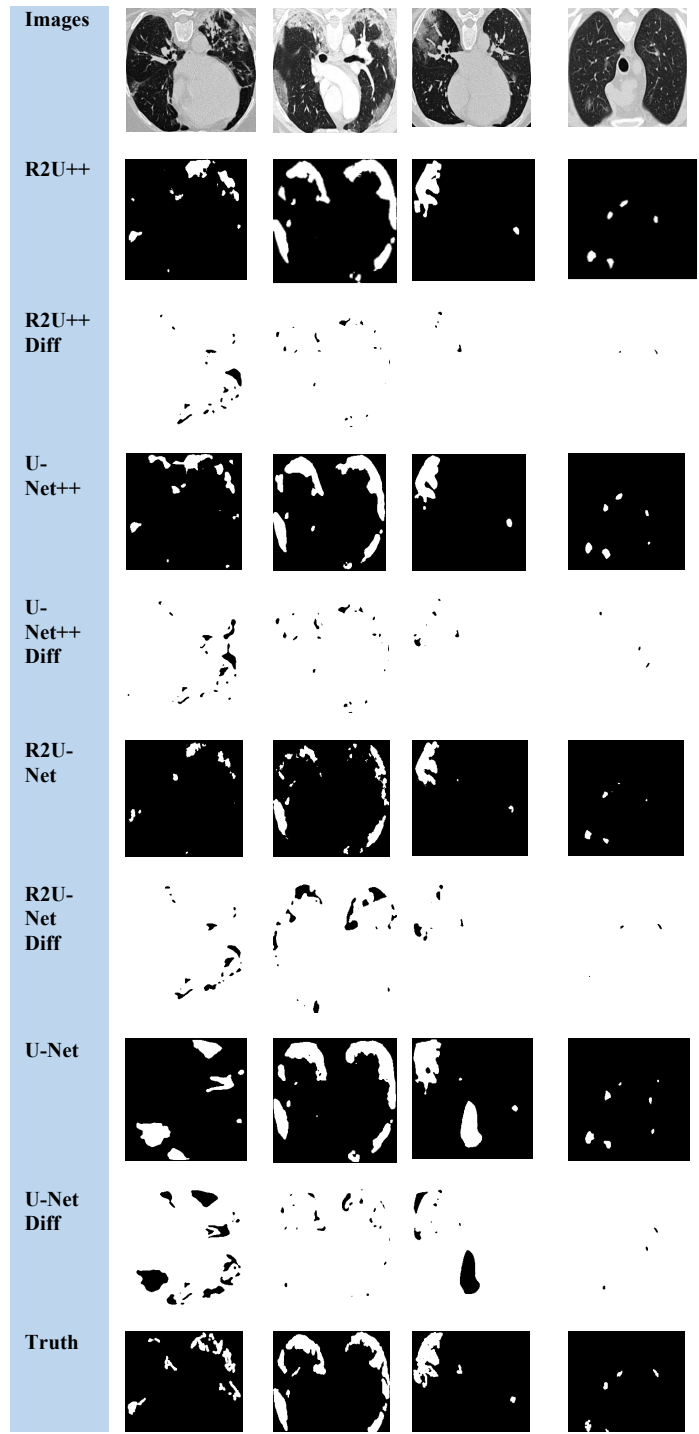

**Fig. 8:** The semantic segmentation outputs and difference images with ground truth for COVID-19 dataset from R2U++ (Ours), U-Net++, R2U-Net, and U-Net. The first row has the input image, and the final row contains the ground truth image. Diff represents the difference.

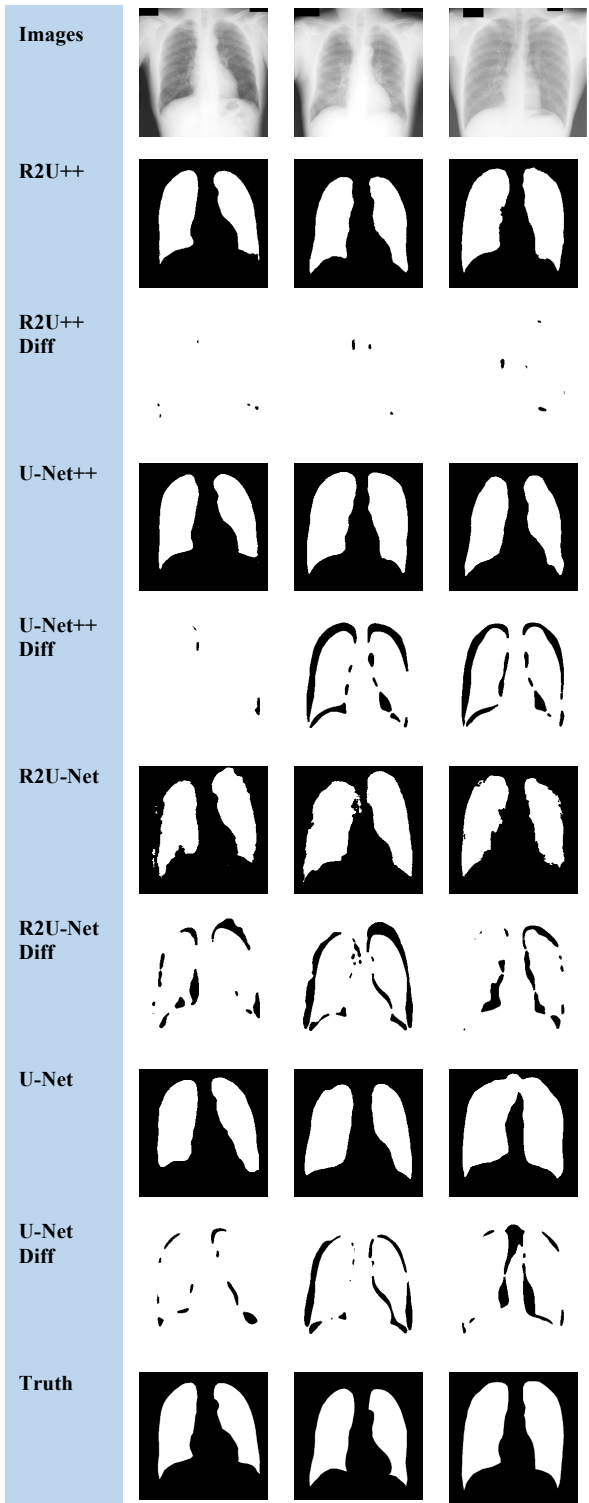

**Fig. 9:** The semantic segmentation outputs and difference images with ground truth for JSRT dataset from R2U++ (Ours), U-Net++, R2U-Net, and U-Net. The first row has the input image, and the final row contains the ground truth image. Diff represents the difference.

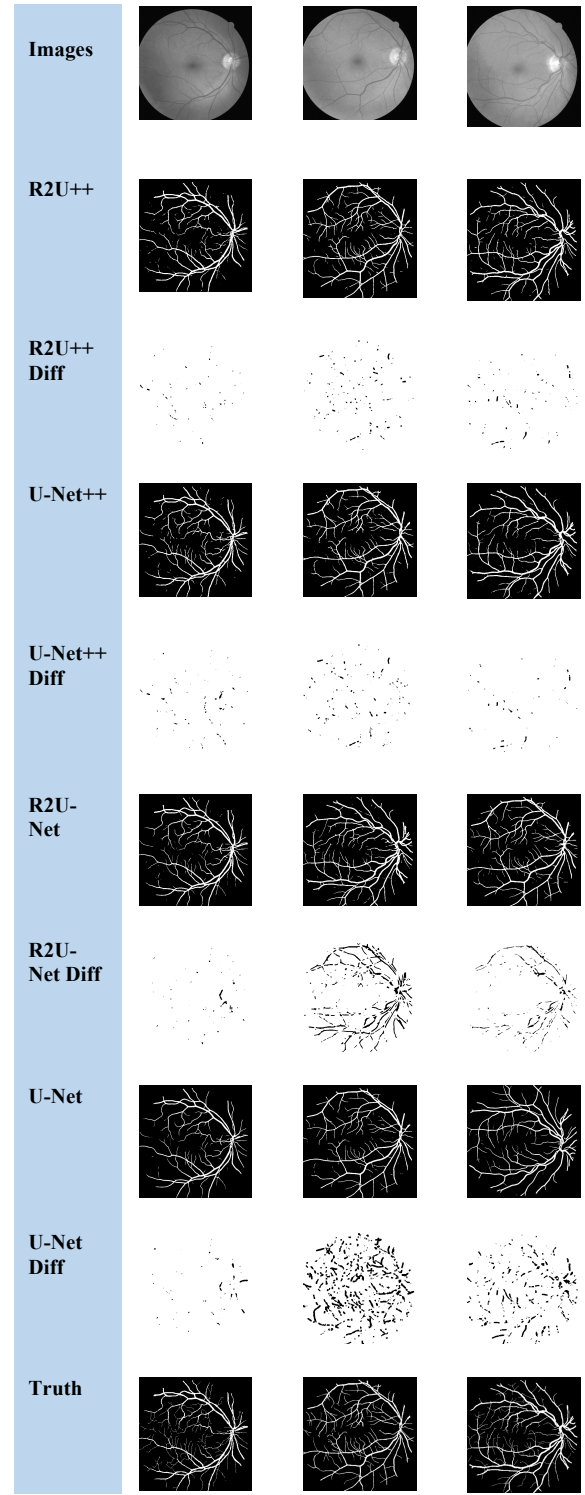

**Fig. 10:** The semantic segmentation outputs and difference images with ground truth for DRIVE dataset from R2U++ (Ours), U-Net++, R2U-Net, and U-Net. The first row has the input image, and the final row contains the ground truth image. Diff represents the difference.

*B. Quantitative Analysis Approaches*

For the analysis of the experimental results the evaluation metrics used in the study are as follows.

*1) Dice coefficient:* The dice coefficient is a commonly used metric for image segmentation which is computed as follows:

$$DC = 2\frac{|GT \cap PR|}{|GT| + |PR|} \quad (8)$$

Where, GT represents the ground truth labels and PR represents the predicted labels.

*2) Accuracy:* Accuracy is used to measure the pixels that are correctly classified by the network. The formula used to calculate accuracy is given by equation:

$$Accuracy = \frac{TP + TN}{TP + TN + FP + FN} \quad (9)$$

Where, TP is true positive, TN is true negative, FP is false positive, and FN is false negative.

*3) Intersection over union:* Another commonly used metric for image segmentation is intersection over union (IoU). It is computed as ratio of intersection of ground truth and predicted results with union of ground truth with predicted labels. The formula is given below:

$$IoU = \frac{|GT \cap PR|}{|GT \cup PR|} \quad (10)$$

*C. Baseline and implementation*

We have compared the performance of our proposed model with U-Net, R2U-Net, and U-Net++. The details of the architecture and number of filters used in the study are shown in Table II. The numbers of filters used in the proposed model are [32, 64,128,256,512]. For each convolution block $X^{m,n}$, the number of filters used are shown in Table II, for example for m=0 and n=0 to 4, i.e. block $X^{0,0-4}$, 32 filters are used. The filter size is kept 3×3 in all layers with a stride of 2. The down-sampling is done using max-pooling operation with a filter size of 2×2 and a stride of 2. The batch normalization is followed by the activation function ReLU. In the final layer sigmoid activation is used to generate predicted probabilities values. We have used Adam optimizer with the learning rate set to 3e-4. All the experiments are implemented using Keras and Tensorflow libraries on NVIDIA GeForce **RTX 2060** with 6 GB dedicated memory. For the training, we have used *early-stop* method on the validation datasets.

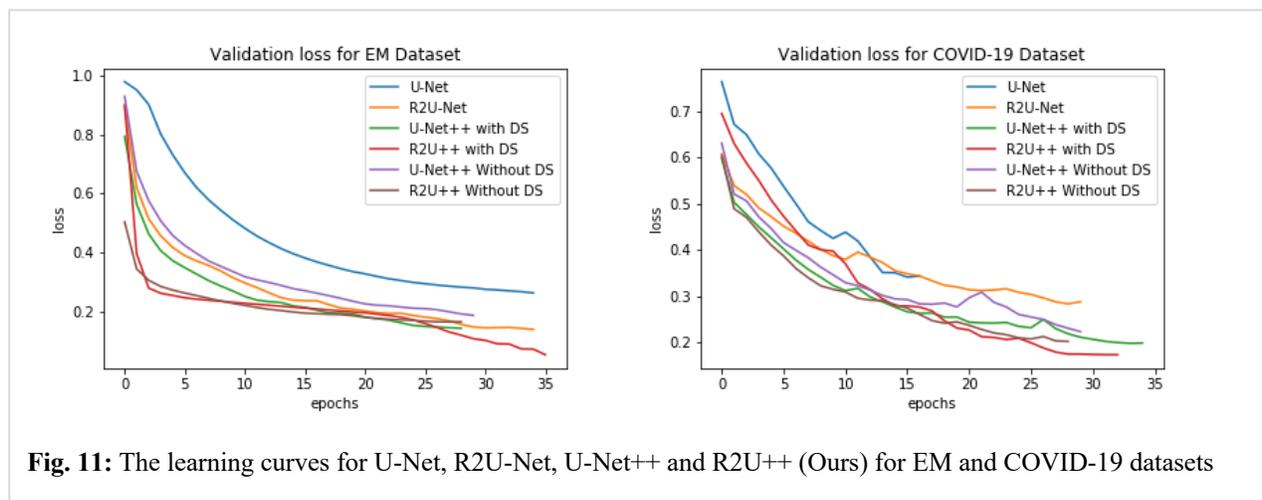

**Fig. 11:** The learning curves for U-Net, R2U-Net, U-Net++ and R2U++ (Ours) for EM and COVID-19 datasets

## V. RESULTS AND DISCUSSION

The results of the R2U++ are compared with the U-Net, R2U-Net, and U-Net++ model in terms of evaluation metrics IoU and dice coefficient for EM, COVID-19 and JSRT dataset. These networks are trained for 20 independent trials and mean IoU and mean dice coefficient with standard deviation (sd) are reported for these trials. The performance on the Drive data set is evaluated using dice coefficient, sensitivity, specificity, and accuracy. The results reported in Table III and Table IV show that R2U++ consistently outperforms U-Net++. In summary, the IoU improvement achieved over UNet++ is up to 3.58% without deep supervision, and up to 1.87% with deep supervision. Similarly, compared with R2U-Net, the IoU improvement is up to 7.11%. The details are as follows.

The improvements in comparison to U-Net++ in terms of mean IoU and dice coefficient without deep supervision are: (2.03↑, 1.61↑) for COVID-19, (3.58↑, 2.11↑) for JSRT, (0.91, 0.05↑) for EM. With deep supervision, the improvements in terms of mean IoU and dice coefficient are: (1.72↑, 1.31↑) for COVID-19, (1.87↑, 1.09↑) for JSRT, (1.02↑, 0.52↑) for EM. Similarly, the improvement over R2U-Net with deep supervision is (4.98↑, 5.05↑) for COVID-19, (7.11↑, 4.55↑) for JSRT, (0.56↑, 0.81↑) for EM. It is evident from the results that adding the recurrent residual connection has shown decent improvement in the performance for both cases.

**Table III:** Segmentation results for EM, COVID-19 and JSRT datasets for U-Net, R2U-Net, U-Net++ and R2U++. The results are reported as mean IoU±sd and Dice±sd on 20 independent trials for both networks with and without deep supervision (DS). Standard deviation is represented in short by sd. The best scores are highlighted.

| Network | Value of $t$ | Parameters | DS | Application |  |  |  |  |  |
|---|---|---|---|---|---|---|---|---|---|
|  |  |  |  | EM |  | COVID-19 |  | JSRT |  |
|  |  |  |  | IoU ± sd | Dice ± sd | IoU ± sd | Dice ± sd | IoU ± sd | Dice ± sd |
| U-Net [14] | - | 7.0M | - | 88.45±1.12 | 93.17±0.64 | 38.29±5.07 | 55.38±5.48 | 76.34±9.46 | 86.22±6.85 |
| R2U-Net [16] | 2 | 16.7M | - | 89.79±0.31 | 94.12±0.17 | 57.82±2.39 | 72.10±2.02 | 81.66±12.59 | 89.50±10.03 |
| U-Net++ [3] | - | 9.0M | ✗ | 88.92±0.14 | 94.09±0.23 | 58.56±1.59 | 73.85±1.29 | 82.17±1.81 | 90.2±1.09 |
| U-Net++ [3] | - | 9.0M | ✓ | 89.33±0.10 | 94.41±0.45 | 61.08±0.04 | 75.84±0.03 | 86.9±3.18 | 92.96±1.88 |
| R2U++ (Ours) | 2 | 18.0M | ✗ | 89.83±0.34 | 94.14±0.19 | 60.59±0.01 | 75.46±0.01 | 85.75±2.09 | 92.31±1.22 |
| R2U++ (Ours) | 2 | 18.0M | ✓ | **90.35±0.23** | **94.93±0.13** | **62.80±0.01** | **77.15±0.01** | **88.77±1.36** | **94.05±0.77** |

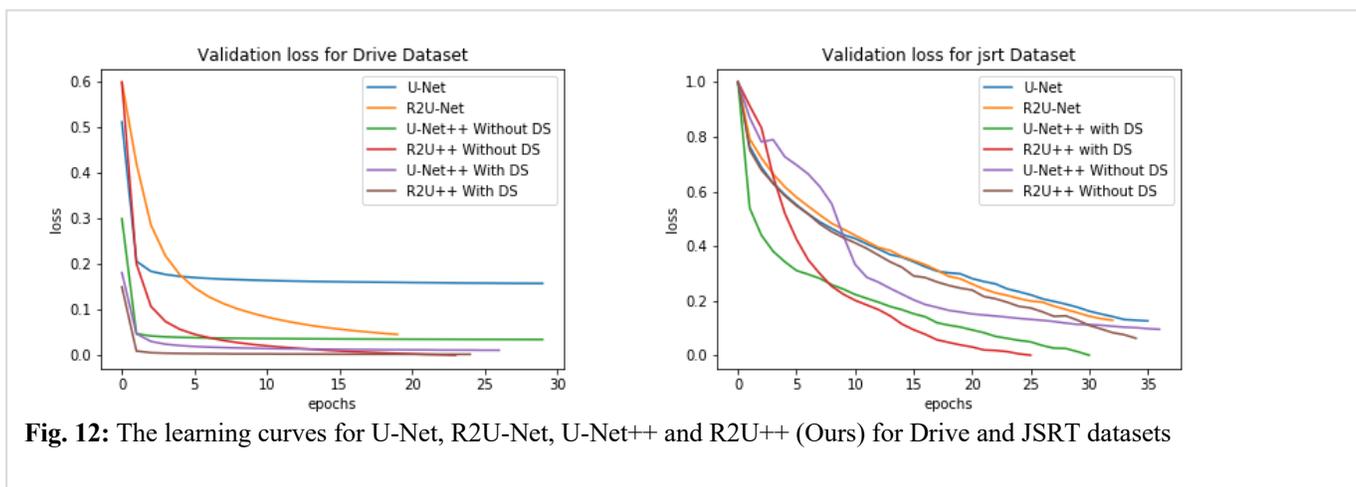

**Fig. 12:** The learning curves for U-Net, R2U-Net, U-Net++ and R2U++ (Ours) for Drive and JSRT datasets

The nature of the complexity of the EM dataset is different than the others because a major part of the image has foreground pixels and very thin blood vessels belong to the background. R2U-Net has more IoU than U-Net++ on EM which shows that recurrent residual connections can help to draw clear boundaries of thin background classes from majority foreground classes. The dice coefficient achieved by our method for COVID-19 is higher than the reported dice coefficient by Inf-Net [46] by a factor of 3.25↑. The segmented images and difference images with ground truths for EM, COVID-19, and JSRT, DRIVE datasets are shown in Fig. 7, Fig. 8, Fig. 9 and Fig 10 respectively. In the case of EM segmentation in Fig. 7, the comparison of row 2, row 4, row 6, and row 8 shows that with R2U++, the contours of cells are segmented properly while preserving the thickness of cell membranes with no breakage. Similarly, for COVID-19, the contours from R2U++ are better defined than U-Net++ which are more rounded as shown in Fig. 8. In addition to this, U-Net++ also has more false positives than R2U++. Similar behavior can be observed for JSRT in Fig. 9.

Experimental results for the DRIVE dataset are reported in Table IV, in comparison with U-Net++ without deep supervision, the increase in dice coefficient, sensitivity, specificity, and accuracy is 0.13↑, 0.02↑, -0.04↓, and 0.1↑ respectively. With deep supervision, the improvement attained in dice coefficient, sensitivity, specificity and accuracy is 0.59↑, 1.18↑, 0.22↑, and 0.09↑, respectively. It can be observed from difference images shown in Fig. 10 that R2U++ is slightly better than U-Net++ in segmenting thin blood vessels. Similarly, the improvement over R2U-Net is 0.64↑ in dice coefficient, 0.82↑ in specificity values, and 0.44↑ in accuracy value with deep supervision.

The learning curves for the datasets by each model are shown in Fig. 11 and Fig. 12 using loss function from equation 7 for no deep supervision and with deep supervision, respectively. It is obvious that R2U++ has the lowest validation error in all the cases. The comparison of inference time taken by models under study is shown in Fig. 13. The models have been tested on 20,000 drive patches with the size of 96×96. As expected, U-Net having the least number of parameters takes the least amount while our model takes the most.

While the proposed method consistently outperformed U-Net++ and U-Net on the segmentation tasks, we observed that there is a significant increase in the number of trainable parameters and thus, an increase in the required computational resources for training the model. However, we believe that this requirement is alleviated by the larger memory and number of cores in modern GPUs that are rapidly becoming available. Furthermore, by modern standards of deep learning, the proposed model with parameters in the order of 18M looks smaller when compared to more recent models such as vision-transformers that have parameters in the order of 632M [49].

**Table IV:** The recorded dice coefficient, sensitivity, specificity and accuracy values for U-NET, R2U-NET, U-NET++ and R2U++. The best scores are highlighted.

| Network | DS | DRIVE dataset | | | |
|---|---|---|---|---|---|
| | | Dice coefficient | Sensitivity | Specificity | Accuracy |
| **U-NET [11]** | - | 79.58 | 74.65 | 98.55 | 96.15 |
| **R2U-Net [16]** | - | 80.86 | **80.87** | 97.86 | 96.16 |
| **U-Net++ [15]** | ✗ | 79.89 | 73.69 | 98.80 | 96.36 |
| **U-Net++ [15]** | ✓ | 80.91 | 76.87 | 98.72 | 96.51 |
| **R2U++ (Ours)** | ✗ | 80.02 | 73.71 | **99.02** | 96.46 |
| **R2U++ (Ours)** | ✓ | **81.50** | 78.05 | 98.68 | **96.60** |

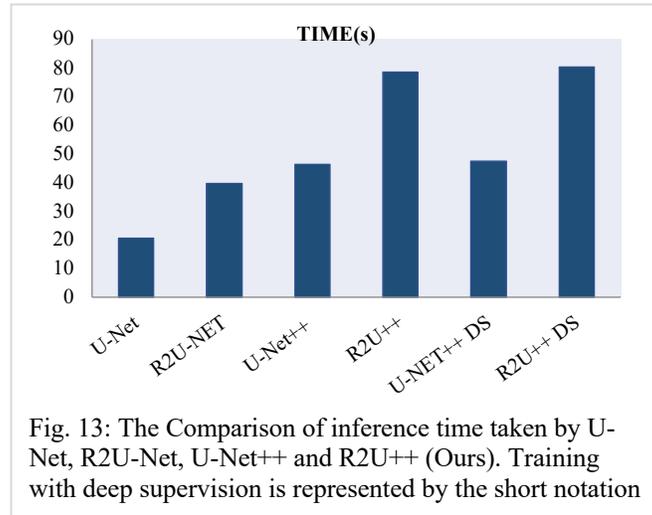

Fig. 13: The Comparison of inference time taken by U-Net, R2U-Net, U-Net++ and R2U++ (Ours). Training with deep supervision is represented by the short notation

## VI. Conclusion

In this study, we introduced recurrent residual convolution blocks and dense skip connections-based U-Net architecture for medical image segmentation. The proposed architecture extracts the features best representing "what" and "where" information, which is backed by the performance of model. The improvement in the performance of the segmentation task can be attributed to; 1) the use of recurrent residual unit over a plain convolution which enables the network to extract low level features precisely without running into the degradation problem, 2) the dense skip pathways help in reducing the semantic gap between encoder and decoder thus more similar semantic concatenation results in improved performance and 3) the deep supervision enables us to classify the multiscale foreground objects correctly. The performance of R2U++ is evaluated on four distinct medical imaging modalities: electron microscopy (EM), X-rays, fundus, and computed tomography (CT). The average gain achieved in IoU score is $1.5\pm0.37\%$, and in dice score is $0.9 \pm 0.33\%$ over UNET++, whereas $4.21\pm2.72$ in IoU, and $3.47\pm1.89$ in dice score over R2U-Net across these different medical imaging segmentation datasets. Our future work will focus on exploring the use of dense skip connections in deep generative models, particularly generative adversarial networks for medical image segmentation.

**Conflict of interest:**
The authors declare no conflict of interest.

17[9] X. Yi, E. Walia and P. Babyn, "Generative adversarial network in medical imaging: A review," *Medical image analysis,* vol. 58, p. 101552., 2019..

[10] J. Long, E. Shelhamer and T. Darrell, "Fully Convolutional Networks for Semantic Segmentation," *IEEE Transactions on Pattern Analysis and Machine Intelligence,* vol. 39, no. 4, pp. 640-651, 2017.

[11] O. Ronneberger, . P. Fischer and T. Brox, "U-Net: Convolutional Networks for Biomedical Image Segmentation," *arXiv:1505.04597v1,* 2015.

[12] N. Ibtehaz and M. S. Rahman, "MultiResUNet: Rethinking the U-Net architecture for multimodal biomedical image segmentation," *Neural Networks,* vol. 121, pp. 74-87, 2019.

[13] Z. Zhou, M. M. R. Siddiquee, N. Tajbakhsh and J. Liang, "UNet++: A Nested U-Net Architecture for Medical Image Segmentation," *arXiv:1807.10165,* 2018.

[14] F. Chen, Y. Ding, Z. Wu, D. Wu and J. Wen, "An Improved Framework called DU++ Applied to brain," in *15th International Computer Conference on Wavelet Active Media Technology and Information Processing (ICCWAMTIP)*, 2018.

[15] Z. Zhou, M. M. R. Siddiquee, N. Tajbakhsh and . J. Liang, "UNet++: Redesigning Skip Connections to Exploit Multiscale Features in Image Segmentation," *Journal of IEEE Transactions on Medical Imaging,* 2019.

[16] M. Z. Alom, . M. Hasan, . C. Yakopcic, T. M. Taha and . V. K. Asari, "Recurrent Residual Convolutional Neural Network based on U-Net (R2U-Net) for Medical Image Segmentation," *arXiv:1802.06955,* 2018.

[17] O. Oktay, J. Schlemper, L. . L. Folgoc, M. Lee, M. Heinrich, K. Misawa, K. Mori, S. McDonagh, . N. . Y. Hammerla, B. Kainz, B. Glocker and D. Rueckert, "Attention U-Net:Learning Where to Look for the Pancreas," *arXiv:1804.03999v3,* 2018.

[18] J. Zhang, Y. Jin, J. Xu, X. Xu and Y. Zhang, "MDU-Net: Multi-scale Densely Connected U-Net for biomedical image segmentation," *arXiv:1812.00352,* 2018.

[19] L. C. Chen, . G. Papandreou, I. Kokkinos, K. Murphy and A. . L. Yuille, "DeepLab: Semantic Image Segmentation with Deep Convolutional Nets, Atrous Convolution,and Fully Connected CRFs," *arXiv:1606.00915v2,* 2017.

[20] V. Badrinarayanan, A. Kendall and R. Cipolla, "SegNet: A Deep Convolutional Encoder-Decoder Architecture for Image Segmentation," *arXiv:1511.00561v3,* 2015.

[21] R. Li, W. Liu, L. Yang, S. Sun, W. Hu , F. Zhang and W. Li, "DeepUNet: A Deep Fully Convolutional Network for Pixel-Level Sea-Land Segmentation," *IEEE Journal of Selected Topics in Applied Earth Observations and Remote Sensing,* vol. 11, no. 11, pp. 3954 - 3962, 2018 .

[22] X. Glorot and Y. Bengio, "Understanding the difficulty of training deep feedforward neural networks," in *International Conference on Artificial Intelligence and Statistics (AISTATS)*, 2010.

[23] S. Ioffe and . C. Szegedy, "Batch Normalization: Accelerating Deep Network Training by Reducing Internal Covariate Shift," *International conference on machine learning*, pp. 448-456. PMLR, 2015.

[24] K. He, X. Zhang, . S. Ren and J. Sun, "Deep Residual Learning for Image Recognition," in *Proceedings of the IEEE conference on computer vision and pattern recognition*, Las Vegas, NV, USA, 2016.

[25] M. Liang and X. Hu, "Recurrent convolutional neural network for object recognition," in *Procedings of IEEE Conference on Computer Vision and Pattern Recognition*, 2015.

[26] B. Kayalıbay, G. Jensen and P. v. d. Smag, "CNN-based Segmentation of Medical Imaging Data," *arXiv:1701.03056v2,* 2017.

[27] A. Soni, R. Koner and V. G. K. Villuri, "M-UNet: Modified U-Net segmentation framework with satellite imagery," in *Proceedings of the Global AI Congress 2019*.

[28] G. Huang, Z. Liu and L. v. d. Maaten, "Densely connected convolutional networks," in *Proceedings of the IEEE Conference on Computer Vision and Pattern Recoginition*, 2017.

[29] Z. Zhang, C. Wu, S. Coleman and D. Kerr, "DENSE-INception U-net for medical image segmentation," *Computer Methods and Programs in Biomedicine,* vol. 192, 2020.

[30] I. Rubasinghe and D. Meedeniya, "Ultrasound Nerve Segmentation Using Deep Probabilistic Programming," *Journal of ICT Research and Applications,* vol. 13, no. 3, pp. 241-256, 2019.

[31] X. Chen, L. Yao and Y. Zhang, "Residual Attention U-Net for automated multi-class segmentation of COVID-19 chest CT images," *arXiv:2004.05645,* 2020.

[32] Q. Yan, B. Wang, . D. Gong, C. Luo, W. Zhao, J. Shen, Q. Shi, S. Jin, L. Zhang and Z. You, "COVID-19 chest CT image segmentation – A deep convolutional neural network solution," *arXiv:2004.10987,* 2020.